\documentclass[twocolumn]{aastex63}
\usepackage{amsmath, mathtools}
\usepackage{courier}
\usepackage{url}

\newcommand{\msun}{M$_\sun$}
\newcommand{\kms}{km~s$^{-1}$}
\newcommand{\vsini}{\ensuremath{v \sin{i}}}
\newcommand{\kepler}{\textit{Kepler}}
\newcommand{\gaia}{\textit{Gaia}}
\newcommand{\deltanu}{$\Delta\nu$}
\newcommand{\numax}{$\nu_{\rm max}$}

\accepted{September 14th, 2023}

\submitjournal{Astrophysical Journal}

\shorttitle{}
\shortauthors{Bufanda et al.}

\begin{document}

\title{Investigating APOKASC Red Giant Stars with Abnormal Carbon to Nitrogen Ratios}

\author[0000-0002-0406-8518]{Erica Bufanda} 
\affiliation{Institute for Astronomy, University of Hawai'i, 2680 Woodlawn Drive, Honolulu, HI 96822, USA}
\author[0000-0002-4818-7885]{Jamie Tayar}
%\altaffiliation{NASA Hubble Fellow}
\affiliation{Department of Astronomy, University of Florida, Bryant Space Science Center, Stadium Road, Gainesville, FL 32611, USA }
\affiliation{Institute for Astronomy, University of Hawai'i, 2680 Woodlawn Drive, Honolulu, HI 96822, USA}
\author[0000-0001-8832-4488]{Daniel Huber}
\affiliation{Institute for Astronomy, University of Hawai'i, 2680 Woodlawn Drive, Honolulu, HI 96822, USA}

\author[0000-0001-5388-0994]{\text{Sten Hasselquist}}
\affiliation{Space Telescope Science Institute, 3700 San Martin Drive, Baltimore, MD 21218, USA}

\author[0000-0003-1805-0316]{Richard R. Lane}
\affiliation{Centro de Investigación en Astronomía, Universidad Bernardo O'Higgins, Avenida Viel 1497, Santiago, Chile}

%\altaffiltext{1}{Institute for Astronomy, University of Hawaii, 2680 Woodlawn Drive, Honolulu, Hawaii 96822, USA}
%\altaffiliation{} 

%\date{\today}

\begin{abstract}
The success of galactic archaeology and the reconstruction of the formation history of our galaxy critically relies on precise ages for large populations of stars. For evolved stars in the red clump and red giant branch, the carbon to nitrogen ratio ([C/N]) has recently been identified as a powerful diagnostic of mass and age that can be applied to stellar samples from spectroscopic surveys such as SDSS/APOGEE. Here, we show that at least 10\% of red clump stars and %$\approx 10\%$ of
red giant branch stars deviate from the standard relationship between [C/N] and mass. 
{We use the APOGEE-\kepler\ (APOKASC) overlap sample to show that binary interactions are %the majority contributors to these
responsible for the majority of these outliers and that stars with %any
indicators of current or previous binarity should be excluded from galactic archaeology analyses that rely on [C/N] abundances to infer stellar masses. We also show that the %standard 
DR14 APOGEE analysis overestimates the surface gravities for even moderately rotating giants (vsini$>2$ km/s)}.
\end{abstract}

\section{Introduction}
Galactic archaeology aims to reconstruct the formation and evolution of the Milky Way and other galaxies. %One method of doing this uses
Critical ingredients to achieve this are ages, chemistry, and kinematics of stars to trace them back to their birth locations \citep{Hogg2016, freeman2002}. 
%However, such analysis requires precise manual inspection for large numbers of stars, which can be difficult to achieve. 
Evolved red giant stars are prime candidates for constructing age maps across the galaxy because they are intrinsically bright. Thus knowledge of their masses and metallicities can be used to infer ages from theoretical evolutionary tracks.
%However,currently the
While metallicities can be measured spectroscopically, masses are more challenging to obtain. The most precise method of obtaining stellar masses of single field stars is asteroseismology, the study of stellar oscillations \citep{KjeldsenBedding1995, Miglio2013}. However, the detection of oscillations requires high-precision and high-cadence photometry for each individual star, and thus is impractical for very large populations \citep{Pinsonneault2018}.  %Thus asteroseismology is impractical for systematically calculating masses and thus ages for a large population of stars.

Alternatively, %the astronomy community has many more spectra for stars through 
spectroscopic surveys such as LAMOST, APOGEE and GALAH \citep{Zhao2012, Holtzman2015, DeSilva2015} observe many more stars than we have asteroseismic detections. With this in mind, a promising alternative to measuring masses of red giant stars directly has been the use of mass-dependent mixing diagnostics measured from stellar spectra. 

As stars ascend onto the red giant branch, the growing surface convection zone dredges up material that has undergone nuclear processing via the CNO cycle. Since both the maximum depth of the surface convection zone and the rate of CNO burning are temperature dependent, it is expected that the ratio of the amount of carbon to the amount of nitrogen on the surface of red giants should correlate with stellar mass \citep{iben1964}. \citet{MasseronGilmore2015} showed that the the observed carbon-to-nitrogen ratio [C/N] was indeed mass dependent, and \citet{Martig2016} and \citet{Ness2016} have used these empirical relationships, which were calibrated using asteroseismic masses, to estimate masses and ages for thousands of giants across the galaxy. 

A limitation of these relationships is that a substantial fraction of red giants deviated from the expected relationship between [C/N] and mass. Non-canonical mixing in the stellar interior may affect [C/N] in evolved stars. For example, Lower [C/N] than expected given a star's mass may be explained by extra mixing which is common in low metallicity stars above the red giant branch bump \citep[e.g.][]{ Gratton2000,Shetrone2019, Masseron2017}. However, stars with higher [C/N] than expected are harder to explain with stellar interior processes. They could theoretically be formed through mass accretion from an unprocessed companion  [i.e. a large planet, or stars that have not undergone the first dredge up.] %An alternative theory is Nitrogen depletion and thus high [C/N] could be connected to the Helium flash. ]\citep{Masseron2017}}.
Alternatively, it has been suggested that the high nitrogen abundances observed in some red clump stars \citep{Masseron2017} could be connected to the helium flash, although such a mechanism is challenging to understand theoretically.

Since the [C/N]-mass relationship is used as a tool to systematically calculate the masses for hundreds of thousands of evolved giant stars {\citep[e.g][]{Ness2016, Ho2017, Mackereth2019}}, it is %necessary to get a better sense of where the outliers are coming from physically.
critical to physically understand the source and prevalence of these outliers.
%Having a way to flag whether a star's [C/N] will
In particular, creating a diagnostic which indicates whether a [C/N] measurement will accurately predict mass is key to applying this method on a very large scale. In this paper we analyze evolved stars that do not follow the typical [C/N]-mass relation using \kepler, APOGEE, and \gaia\ data. Our goal is to provide recommendations on the constraints of using this relation to measure mass and age for large stellar populations.
 
\section{Observations}

\subsection{APOGEE Data}

Our main data set is the APOGEE-\kepler-2 catalog \citep[APOKASC2]{Pinsonneault2018}, a large stellar catalog of over 6000 giant stars with stellar properties and evolutionary states derived from APOGEE spectroscopic parameters \citep{elsworth2017, Holtzman2018} and %\kepler
asteroseismic data. The APOKASC-2 sample uses spectroscopic parameters and uncertainties taken from the fourteenth data release (hereafter DR14, \cite{DR14}) of the Sloan Digital Sky Survey (SDSS) \citep{Eisenstein2011} from the Apache Point Observatory Galactic Experiment (APOGEE) \citep{Majewski2017} which were obtained during SDSS-IV \citep{Blanton2017} operating on the Sloan 2.5 meter telescope \citep{Gunn2006}. APOGEE has acquired over half a million high (R $\sim$ 22,500) resolution infrared spectra \citep{Wilson2012, Wilson2019}.

Updated spectroscopic parameters for these stars are now available from the more recent Data Release 17 (DR17) \citep{DR17}, which includes improvements to the spectroscopic pipeline \citep{Jonsson2020} including updated line-lists \citep{Smith2021}, improved atmospheric models, and so forth. For consistency with the asteroseismic analysis, we continue to use %Data Release 14
DR14 spectroscopic parameters for this work, although initial investigations indicate that the majority of our outliers are still anomalous in DR17, and the overall fraction of outliers is likely to be roughly comparable.

The APOGEE Stellar Parameters and Abundances Pipeline \citep[ASPCAP]{Nidever2015, GarciaPerez2016} derives stellar parameters through a global chi-squared minimization to the entire spectrum and then individual chemical abundances are derived using windows around the relevant lines for each species. For the DR14 catalog, chemical abundances are measured for C, N, O, Na, Mg, Al, Si, S, K, Ca, Ti, V, Mn, Fe, Co, and Ni. These initial results are then calibrated using data from stars of known parameters, including asteroseismic stars, open clusters, and low extinction fields \citep{Holtzman2018}.
%Note: elements are in order of periodic table number

\subsection{{\kepler} Light Curves}

We use the masses and other seismically derived properties reported in the APOKASC-2 catalog, which were measured from analyzing light curves from the \kepler\ mission \citep{Borucki2010, Gilliland2010}. 

For the APOKASC-2 sample, members of the Kepler Asteroseismic Science Consortium analyzed the \kepler\ long cadence photometry using five independent methods (known in the literature as A2Z, CAN, COR, OCT, and SYD) to estimate {\numax} and {\deltanu}
  \citep{Garcia2011, HandbergLund2014} [see \cite{Serenelli2017} for a detailed overview of these methods.]
  %,elsworth2017}.
Systematics between the analysis pipelines were then corrected, and theoretical corrections to \deltanu\ were applied. The results were then put on an empirical scale using the results from open clusters in the \kepler\ field \citep{Pinsonneault2018}. It is worth noting that the corrections applied were different for core helium burning (red clump) stars and shell hydrogen burning (RGB) stars, and that evolutionary state was determined directly from the asteroseismic analysis \citep{Elsworth2019} 

\section{Outlier Diagnostics} \label{sec:outlierdiagnostics}

\subsection{Identifying Outliers}

 \begin{figure}
\begin{center}
\includegraphics[width=8.5cm]{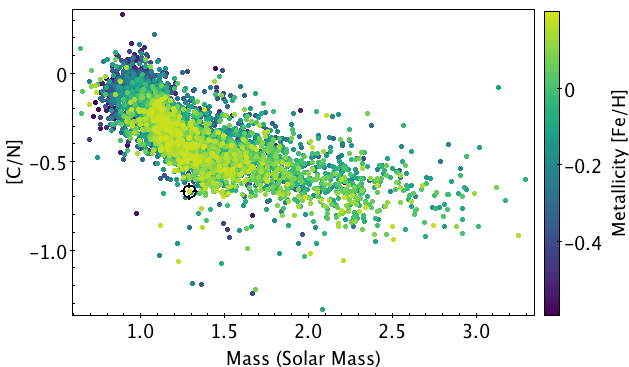}
\caption{The [C/N]-mass relation color-coded by metallicity [Fe/H] for evolved stars in the APOKASC sample. Due to the slight dependence of the slope of the relationship on metallicity, we divide the data into low metallicity ($-0.6<$[Fe/H]$<-0.1$) and high-metallicity ([Fe/H]$>-0.1$) samples. } 
\label{fig:states}
\end{center}
\end{figure}

\begin{figure*}
\begin{center}
\includegraphics[width=17cm]{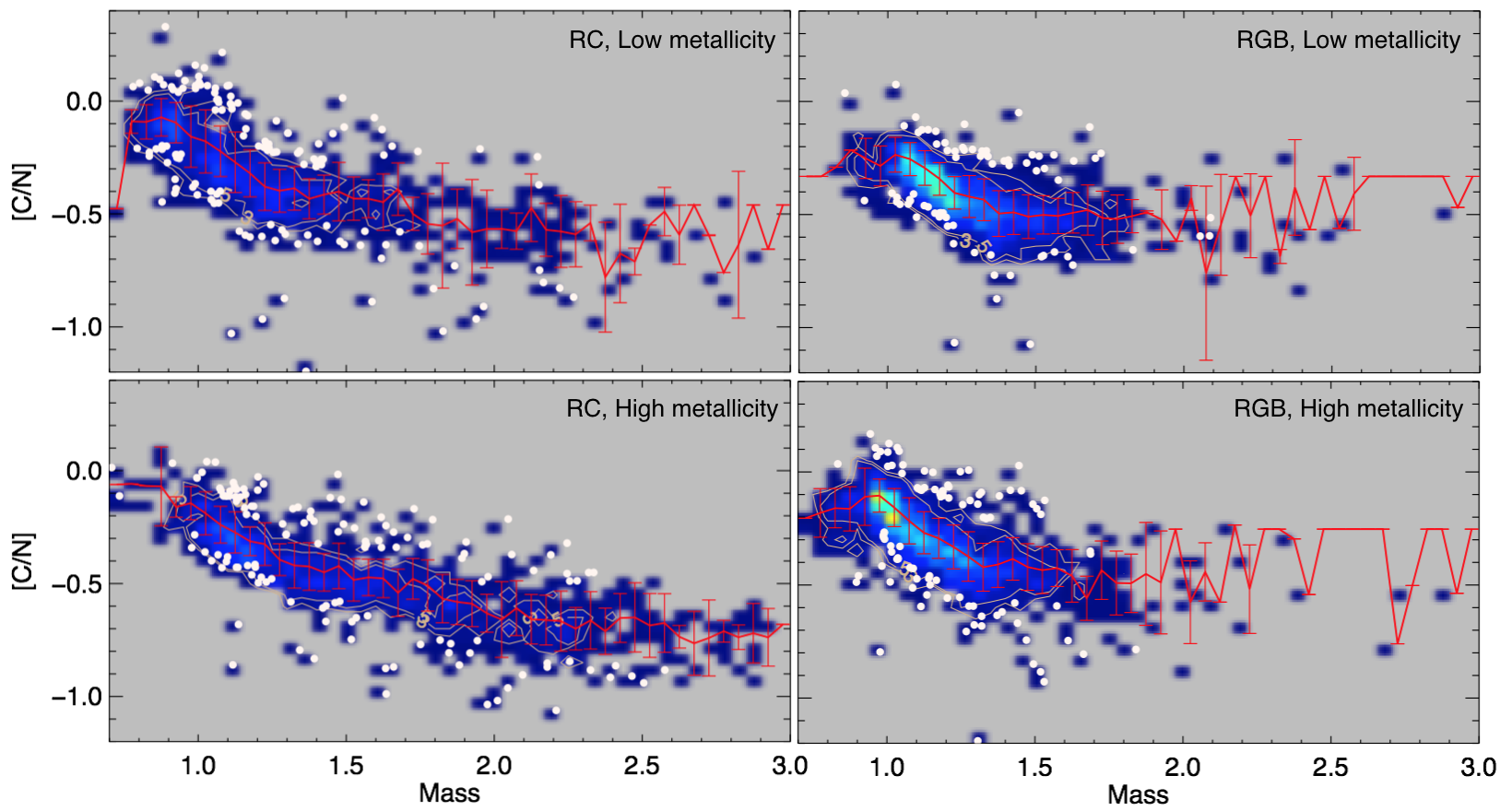}
\caption{A 2D histogram binned in 0.05 solar masses and 0.05 dex in [C/N] displaying the population density of carbon-to-nitrogen as a function of mass for our different populations, in bins of metallicity (top:low metallicity, bottom: high metallicity) and evolutionary state (left:RC, right:RGB). Red points represent the average and standard deviation in each bin; the red trendline is a moving average. Outliers are marked as white points and are at least 1.5$\sigma$ away from the bin average. Because of small numbers, we have excluded from our analysis RGB stars above 2.0\msun\ and RC stars above 2.5\msun.}
\label{fig:c2nmasspop}
\end{center}
\end{figure*}

Figure \ref{fig:states} shows the relation between [C/N] and asteroseismic mass for the stars in our sample. We expect that the mixing as a function of mass should depend on metallicity and evolutionary state. Our initial investigations confirmed that the [C/N]-mass trends were indeed weakly metallicity sensitive, and significantly dependent on evolutionary phase. %We therefore separated the sample into two metallicity bins (above and below [M/H]$=-0.1$), and 
We therefore first separated the sample into two bins based on their asteroseismic evolutionary state: core helium burning, including stars labeled as primary or secondary clump (also known as `red clump' [RC] stars), and those in shell burning phases, both those marked as first ascent red giants (RGB) and those whose energy source is ambiguous (RGB/AGB). We removed stars with unidentified or ambiguous evolutionary states from our analysis.

Due to the limited number of stars with very low ($\leq-0.6$) and high metallicities ($\geq0.2$) in our sample, we exclude these stars from our analysis. Based on this and the weak evolution of the [C/N]-mass slope as a function of metallicity (see Figure \ref{fig:states}) we divide our sample into one `low' (-0.6 to -0.1) and `high' metallicity (-0.1 to 0.2) bin. This yields $2222$ RC and $3166$ RGB stars in our final sample, with 2564 and 2615 low and high metallicity giants respectively. The resulting [C/N] relationship is shown for each of the populations in Figure \ref{fig:c2nmasspop}.

To separate outliers from the [C/N] trend for each of our populations we binned the stars in terms of mass in steps of 0.05 solar masses from 0.7 to 3 solar masses. We calculated the average in each bin, and defined outliers as 1.5 sigma away from the calculated average. %We bin the mass in steps of 0.05 solar masses from 0.7 to 3 solar masses. 
The four resulting outlier populations are shown in Figure 2. The [C/N]-mass relation flattens above 2.0 solar masses, so we concentrate on the mass regime 0.7-2.0 solar masses for this paper.

\subsection{Known Binaries}
We %next 
explore different diagnostics %that may %correlate with
%cause a star %becoming
as a cause for a star to become an outlier in the [C/N] - stellar mass diagram. We first flag binary stars in our outlier sample because binarity may affect the measurement of spectral parameters and chemical abundances. Line blending in a double lined spectroscopic binary, for example, could impact the inferred temperature, gravity, abundances and rotation rates. It is also possible that binary evolution could impact the expected chemical evolution, either through the accretion of unprocessed material, or by driving additional mixing \citep[e.g.][]{Casey2019}. 
%\subsubsection{Eclipsing Binaries}
For example, \citet{Jofre2016, Jofre2023} show that seemingly young and massive stars show evidence of binarity through radial-velocity analyses and do not follow the expected [C/N]-mass trend. %mass tranfer in binary systems can reproduce  abnormal [C/N] and is shown in \citet{Jofre2016}. }
We flag stars that are known eclipsing binaries \citep{Slawson2011}, although most known binaries may have already been eliminated from the APOKASC-2 sample \citep{Tayar2015}.

We additionally flag binaries identified using radial velocity variations, known as ``vscatters'' from multi-epoch APOGEE spectroscopy. The parameter vscatter %is a measurement by APOGEE that 
 measures the maximum radial velocity difference between %measurements
 epochs that have been taken %over
 divided by the square root of the number of observations. High vscatter is shown to correlate with high radial-velocity measurements, and thus can be used as a flag for evidence of binarity \citep{Jofre2023}.

The detection limit of the APOGEE spectrograph and the expected radial velocity jitter for evolved stars are both around 0.5 \kms\ \citep{Badenes2018}. Hence we consider a vscatter above 1 \kms\ as a significant detection of a multiple star system. While the number of observations for each star is not enough to derive an orbital solution for a companion, the variability of radial velocities between observations is sometimes enough to reveal a companion \citep{Badenes2018}.

  \begin{figure*}[t!]
\begin{center}
\includegraphics[width=18cm]{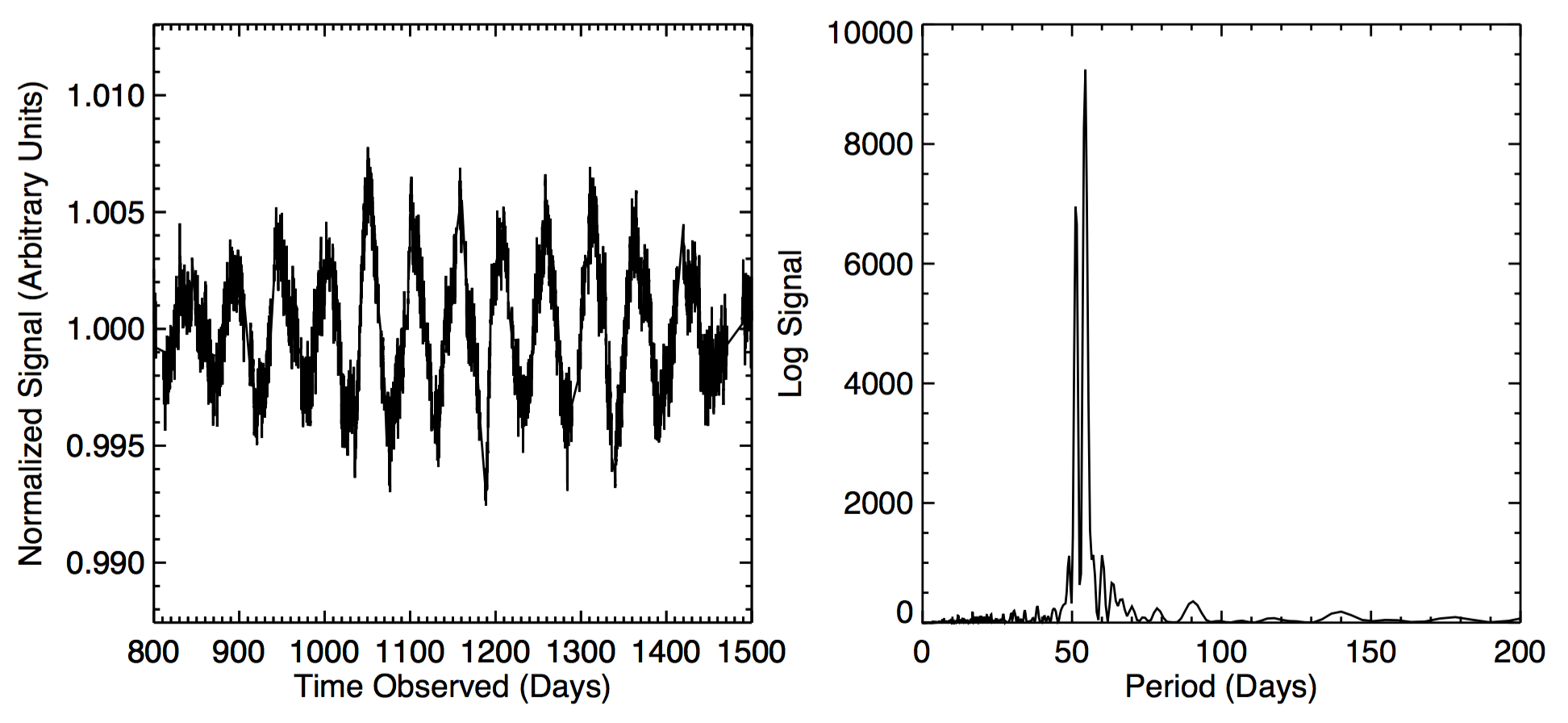}
%\resizebox{.5\width}{!}{\input{logg_plot.png}}
%\includegraphics[width=b6cm]{logg_plot.png}
\caption{Left: The \kepler\ light curve for KIC 11808481  after subtracting long-period sinuisodal signatures ($>200$ days) that are likely due to telescope systematics. The photospheric modulation of the light curve from spots is clear, with a rotation period of 54 days measured from the Lomb-Scargle periodogram (right).
}
\label{fig:sunspots}
\end{center}
\end{figure*}

We note that an insignificant or low vscatter does not necessarily mean the star is not a binary, as the max difference of radial velocities between observations may not reflect the true maximum amplitude of the velocity of the star.
To partially combat this bias, \citet{PriceWhelan2020} has identified a sample of binary candidates in APOGEE through a more sophisticated Monte Carlo search of the available radial velocity measurements. We match their data with stars in our sample to identify 28 additional binaries in the sample. 

Lastly, \citet{Berger2020a} has calculated the Renormalised Unit Weight Error (RUWE) for each star in the \kepler\ field using information from \gaia. RUWE $>$ 1.2 indicates the presence of a close ($<$ 1'') binary by deteriorates the astrometric solution from Gaia \citep{Evans2018}. We thus flag additional binaries in our sample that have \gaia\ RUWE $>$ 1.2.

\subsection{Asteroseismic Measurement Bias}

Stars may also deviate from the observed empirical [C/N]-mass trend is because of ill-defined asteroseismic parameters. In particular, \citet{Pinsonneault2018} noted that high-luminosity red giants (numax $<$ 10 $\mu$Hz) only oscillate in a small number of radial orders. This makes measurements of traditional asteroseismic observables difficult \citep{Stello2014} and confuses %overlapping
mapping from {\numax} and {\deltanu} to stellar parameters. We therefore flag all such stars in our analysis.

\subsection{Rotation}

\subsubsection{Spectroscopic Rotation Velocities} \label{sec:vsini}

Because of angular momentum conservation, the large radii of red giant stars suggest that most stars in this regime should be rotating slowly  \citep[\vsini$<$1 \kms;][]{Tayar2015}
Additionally, for core-helium burning stars, helium flashes have the potential the slow surface rates even further \citep{SillsPinsonneault2000}. At these slow rotation rates, the effect of rotational broadening on spectral lines is rendered unmeasurable by the significantly larger microturbulent and macroturbulent broadening in these stars. Hence detectable rotational broadening for evolved stars may be indicative of additional evolutionary or physical processes such as merger or accretion events, as these have the ability to cause the spin rate of the star to increase \citep[e.g.][]{Patton2023}. %in addition to their potential effects on the surface chemistry. 

%\citet{Tayar2015}  measured \vsini from rotational broadening of the lines in the APOGEE spectra. 
\cite{Tayar2015} measured {\vsini}s for the APOGEE DR10 sample through a cross correlation with broadened versions of the APOGEE template spectra. We perform a similar analysis on the DR14 data, reporting detections for \vsini$>$5 \kms, and flagging as potentially rotating any stars with inferred rotation velocities between 2 and 5 \kms.

\subsubsection{Rotational Modulation}
Rotational modulation due to starspots causes brightness variations that can be used to infer rotation periods. To determine photometric rotation periods we analyzed the Simple Aperture Photometry (SAP) Kepler light curves for our sample \citep{jenkins2010}. We apply a high-pass filter of 200 days to % the \kepler\ light curves to preserve the star-spot signal, correcting
remove systematic trends while preparing star-spot signals and correcting for quarter gaps and other discontinuities with a simple linear fit. %to connect each epoch where we have data.

Given the correlation between rotation rate and star-spot activity \citep{Noyes1984, MamajekHillenbrand2008} we expect the rapidly rotating stars will be more likely to show star-spot modulation. We visually inspect the 66 stars with measurable rotation velocities. We estimated these stars' rotation periods with a Lomb-Scargle periodogram, and estimated the error by fitting a Gaussian to the significant peak (Figure \ref{fig:sunspots}). The results are shown in Table \ref{tab:rotation}.

\begin{figure}
\begin{center}
\includegraphics[width=8.5cm]{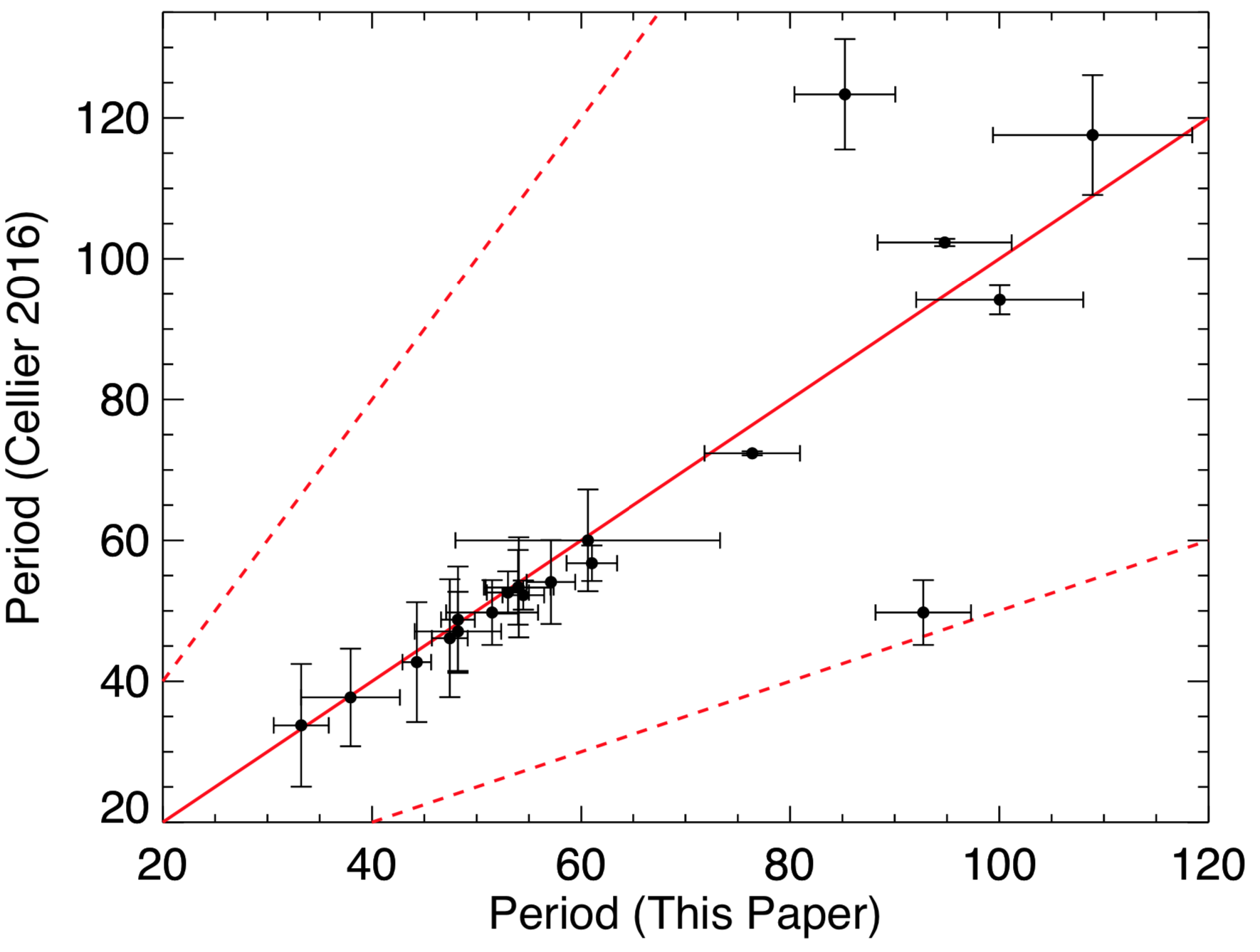}
\caption{Our rotation periods in comparison to  \citet{Ceillier2017}. In red is the 1:1 line. Dashed lines show 1:2 and 2:1 ratios.} 
\label{fig:periodcompare}
\end{center}
\end{figure}

%We demonstrate in 
Figure \ref{fig:periodcompare} compares the values obtained from our method %are consistent
with literature results  %from a more sophisticated analysis that relied on both 
obtained using auto-correlation and wavelet analysis \citep{Ceillier2017}. We removed from our sample 6 stars that were flagged in the \citet{Ceillier2017} analysis as likely to be contaminated. Every star in our velocity-broadened 
%outlier
sample that overlapped with Cellier (2017) had comparable measured rotation and detectable star-spots (Table \ref{tab:rotation}). For 20 stars in common, we find that 18 stars (90 percent) agree to better than 1-$\sigma$. We therefore expect that the additional 12 stars identified by our analysis as new spotted stars are likely real detections of stellar rotation. There are two exceptions; one where we measured twice %or half 
the rotation period as \citet{Ceillier2017}, a common systematic challenge in star-spot modulation measurements \textbf{\citep[see][]{Aigrain2015}}. One target in our sample does not match the value calculated by \citet{Ceillier2017} or the 2:1 or 1:2 of the rotation. We currently can not explain the discrepancy in measured rotation for this star.

More recently, \cite{gaulme2020} searched a subset of the Kepler stars for periodic rotation signals. We have three stars that overlap with that sample and our periods are in good agreement with the published rotation periods (Table \ref{tab:rotation}). We also ensured that all our measured rotation rates are larger than the critical period for which a star would be ripped apart by the centrifugal force \citep{Ceillier2017}: %This period is 

\begin{equation}
T_{crit} = \sqrt{\frac{27\pi^{2}R^{3}}{2GM}}
\end{equation}

Where R and M are the radius and the mass of the star respectively. In general, the critical rate for our stars is between 7 to 10 days, which was not significantly close to any of our measured rotation periods.% of any of the stars in our sample.

\subsection{Chemical Anomalies} \label{ssec:chemAnom}
 \begin{figure}
\begin{center}
\includegraphics[width=7.5cm]{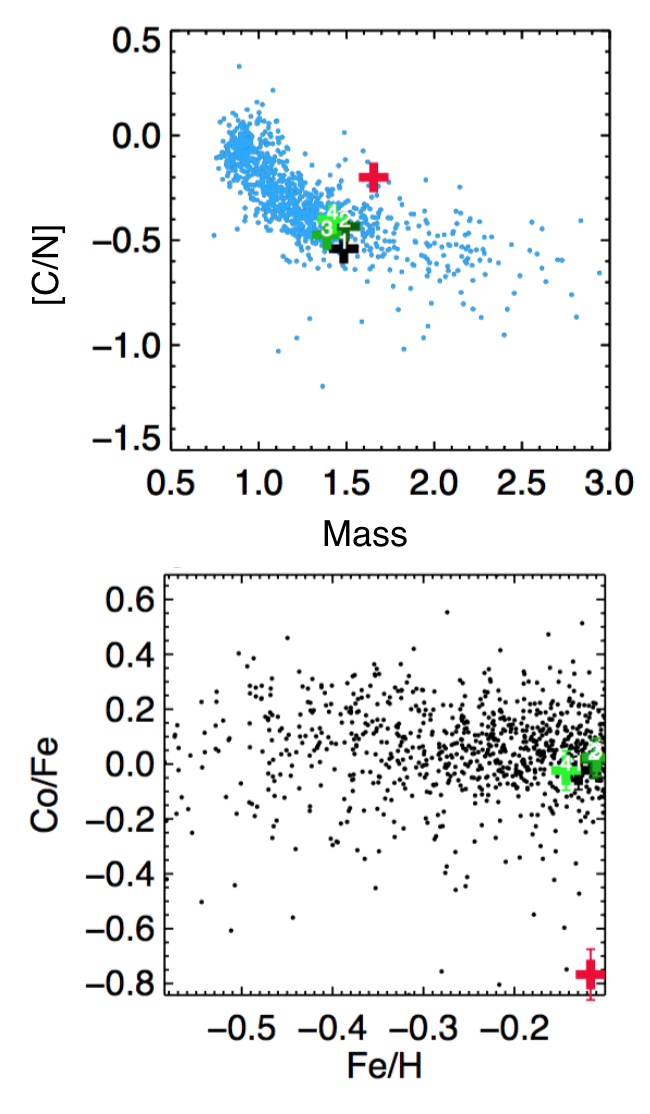}
\caption{
Carbon-to-Nitrogen abundances versus stellar mass (top panel) and Cobalt-to-Iron abundances versus metallicity (bottom panel) for our sample. Colored crosses mark the same stars in both panels. The red cross marks an outlier star (KIC 3735699) which is depleted in Cobalt compared to the rest of the sample.}
\label{fig:matchedpairs}
\end{center}
\end{figure}

In addition to looking at the carbon and nitrogen abundances, we check for offsets in other abundances that could represent either data processing problems or some unusual event. In particular, \citet{Weinberg2019} found that the individual abundances derived by APOGEE can be well predicted using just the [Fe/H] and [$\alpha$/Fe] measurements. This is because these values are intimately related to the ratio enrichment from core collapse supernova to Type Ia supernova, which depends on a star's formation environment and age. We therefore suggest that individual elemental abundances that deviate significantly from the expectation given [Fe/H] and [$\alpha$/Fe] could be the result of contamination from a binary companion \citep[e.g. barium stars,][]{McClure1980} or due to some unusual mixing event. We show this analysis in Figure \ref{fig:matchedpairs}, where the [C/N] vs mass (top) and [Co/Fe] vs [Fe/H] for the general population (blue, top; black, bottom) and for a star from our outlier sample, KIC3735699 (red cross). The green crosses represent four stars that do follow the [C/N]-mass trend that have similar stellar parameters as KIC3735699, and were chosen by minimizing the combined weighted difference (or chi-square statistic) of metallicity, alpha abundance, surface gravity and temperature. We include surface gravity and temperature as a criteria for star-matching because these parameters have a strong influence on %the strength and appearance of the 
spectral lines, and thus could impact the inferred chemical abundances. In comparison, KIC3735699 has a significantly lower [Co/Fe] abundance in comparison to the four matched  stars. 

While we see no obvious evidence that either the goodness of fit metric or radial velocities of the chemically anomalous stars are correlated in a way that would indicate mechanical issues with the spectroscopic analysis e.g. improperly subtracted skylines, further investigations are encouraged to determine if these trends are astrophysical, or related to correlated issues in the abundance determinations. We flag stars that have greater than 3-$\sigma$ offsets in any elemental abundances measured by APOGEE compared to the four matched stars that follow the [C/N]-mass trend (For example, Figure \ref{fig:matchedpairs}, bottom panel). 

\subsection{Alpha/Age Inconsistencies}

 \begin{figure}
\begin{center}
\includegraphics[width=8.5cm]{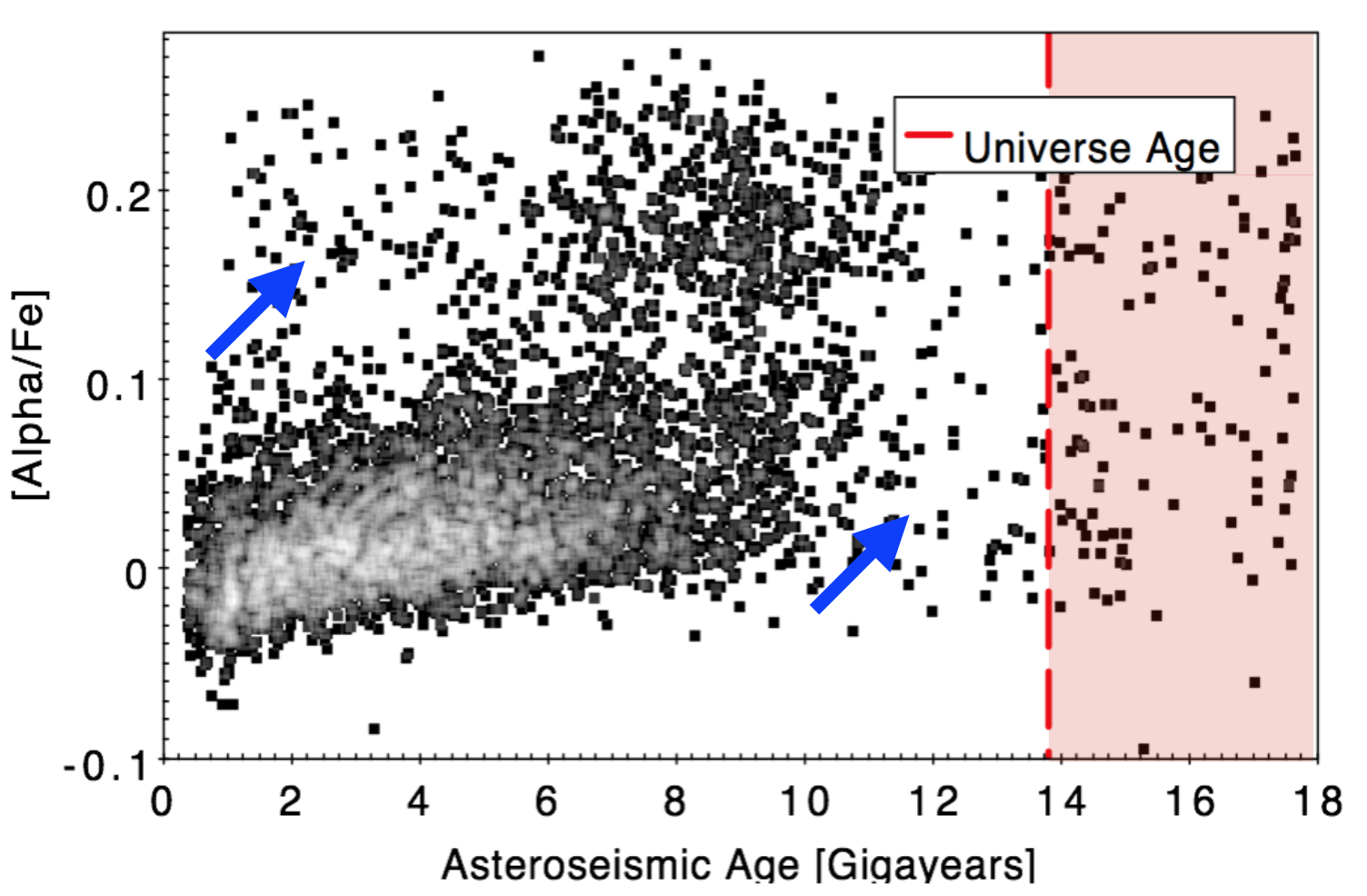}
\caption{
[$\alpha$/Fe] as a function of age for the APOKASC sample. Ages were calculated from the asteroseismic data by \citet{Pinsonneault2018}. The $\alpha$-rich and $\alpha$-poor galactic populations are both visible, as is the slight evolution in $\alpha$-element abundance with time in the thin disk population. %There are regions outside of this typical evolution that are indicated by arrows. 
We have marked anomalous stars that fall significantly outside this standard trend, shown by the blue arrows, including young $\alpha$-rich stars \citep[e.g.][]{Martig2015}, as well as old $\alpha$-poor stars. We have also flagged stars with quoted ages older than the age of the universe (region to the right of the red line).} 
\label{fig:alphaage}
\end{center}
\end{figure}

Figure \ref{fig:alphaage} shows [$\alpha/$Fe] versus asteroseismic age for the APOKASC sample. For most stars we observe a strong correlation between $\alpha$-element abundances and age, which originates from galactic chemical evolution \citep{Fuhrmann1998}. There are two regions that have %potentially unusual
stars whose $\alpha$-element abundances do not follow these age trends (Figure \ref{fig:alphaage}).One of the arrows points to a region that includes the so-called ``young" $\alpha$-rich population ($\alpha>0.1$, age $<$ 6 gigayears), identified as potentially the result of stellar mergers \citep{Martig2015, Sun2020} or as stars that have migrated from an unusual part of the inner galaxy \citep{Chiappini2015}. The other arrow points to the old $\alpha$-poor stars ($\alpha<0.1$, age $>$ 10 gigayears) which are also significantly offset from the normal chemical evolution trend. We flag these stars in our sample.

Next, to the right of the red dashed line in Figure \ref{fig:alphaage} are stars whose age estimates from \citet{Pinsonneault2018} are older than the age of the universe (13.8 Gyr). While some of these stars may but pushed to older ages by random or systematic uncertainties in their age calculations, we suspect that many of these stars have lost mass, and are therefore being inferred to be older than they truly are \citep{Li2022}. This may be because the models used to calculate the ages do not include mass loss, or are not correct for these stars. We also flag these stars in our sample.

\section{Results}

\subsection{Impact of APOGEE log(g) offset on \vsini}

\begin{figure*}
\begin{center}
\includegraphics[width=18cm]{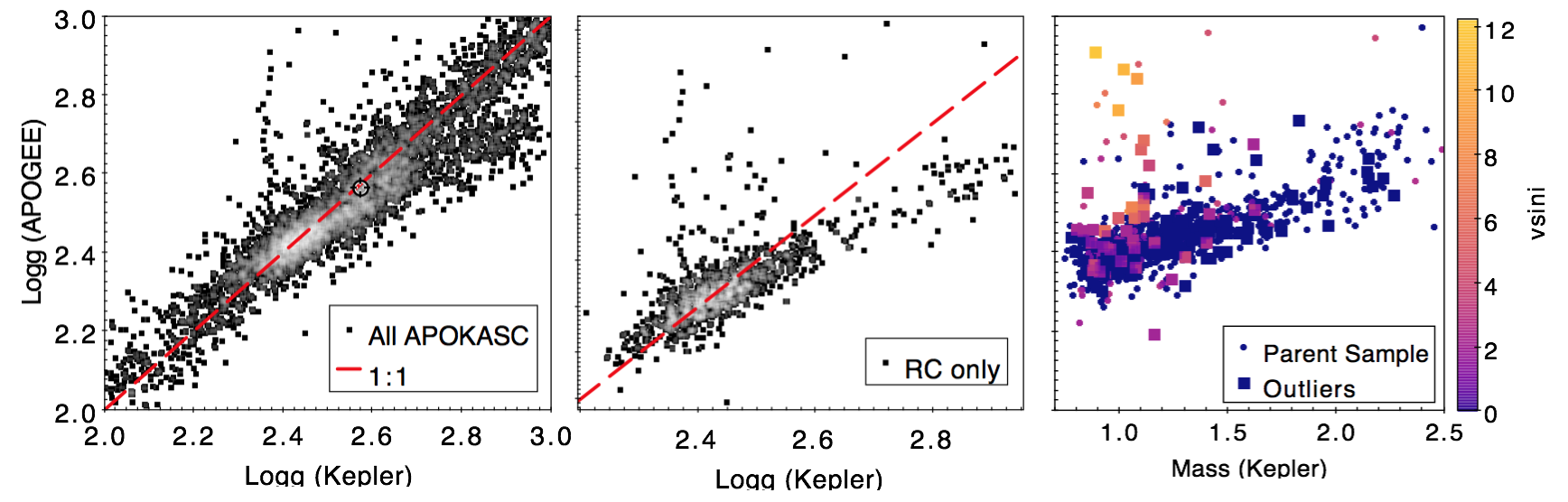}
\caption{Left: Surface gravity measured by APOGEE vs Surface gravity measured by Kepler through asteroseismology for the APOKASC sample (black-grey density, light grey = most dense). The red line shows a 1:1 correspondance. Center: Same axes and 1:1 trend for only the red clump stars in the APOKASC sample. Right: Surface gravity measured by APOGEE vs Mass measured by Kepler for the Red clump sample, differentiating between stars that follow the [C/N]-mass trend (``parent sample'', circles) and stars from the outlier sample (boxes). The stars are color coded by vsini.} 
\label{fig:filter1}
\end{center}
\end{figure*}

APOGEE preforms a simultaneous chi-squared minimization to estimate the temperature, gravity, and bulk composition. Thus, offsets in one parameter could produce incorrect measurements of other properties inferred from the spectra, including the measured [C/N] ratio. Furthermore, an error in the calculation of %surface gravity
the base spectrum can also affect the way we measure rotation from spectra.  Generally speaking, the projected rotation velocity (vsini) is estimated by broadening the template fit to the APOGEE spectrum. If the lines in this spectrum are over-broadened, for example, because the assumed gravity is too high, then we would expect to underestimate the rotational broadening.

In general, the asteroseismic and spectroscopic estimates of surface gravity agree extremely well (see Fig. 7, left) because the spectroscopic results are quite precise and are calibrated on the asteroseismology \citep{Holtzman2018, Jonsson2020}. However there are two issues specifically associated with red clump stars. The first is that all of the APOGEE surface gravities of these stars are systematically offset from their asteroseismic counter-parts (Figure 7, middle). While there are offsets in the raw spectroscopic results \citep{MasseronHawkins2017}, these are supposed to be corrected by a calibration to the asteroseismic scale \citep{Holtzman2018}. However, in this particular data release, there were some errors in that calibration that cause the systematic offset seen in the middle panel. More interesting from our perspective however, is that even after accounting for this offset, we find an additional offset that seems to be present mostly in rapidly rotating stars (Fig. 7, right).

The measured spectroscopic gravity of the clump is strongly correlated with stellar mass, and significant outliers from this trend are often rapidly rotating outliers from the [C/N]-mass trend, suggesting correlated errors between the spectroscopic parameters when rotation is not included in the fits.

\begin{figure}
\begin{center}
\includegraphics[width=8.5cm]{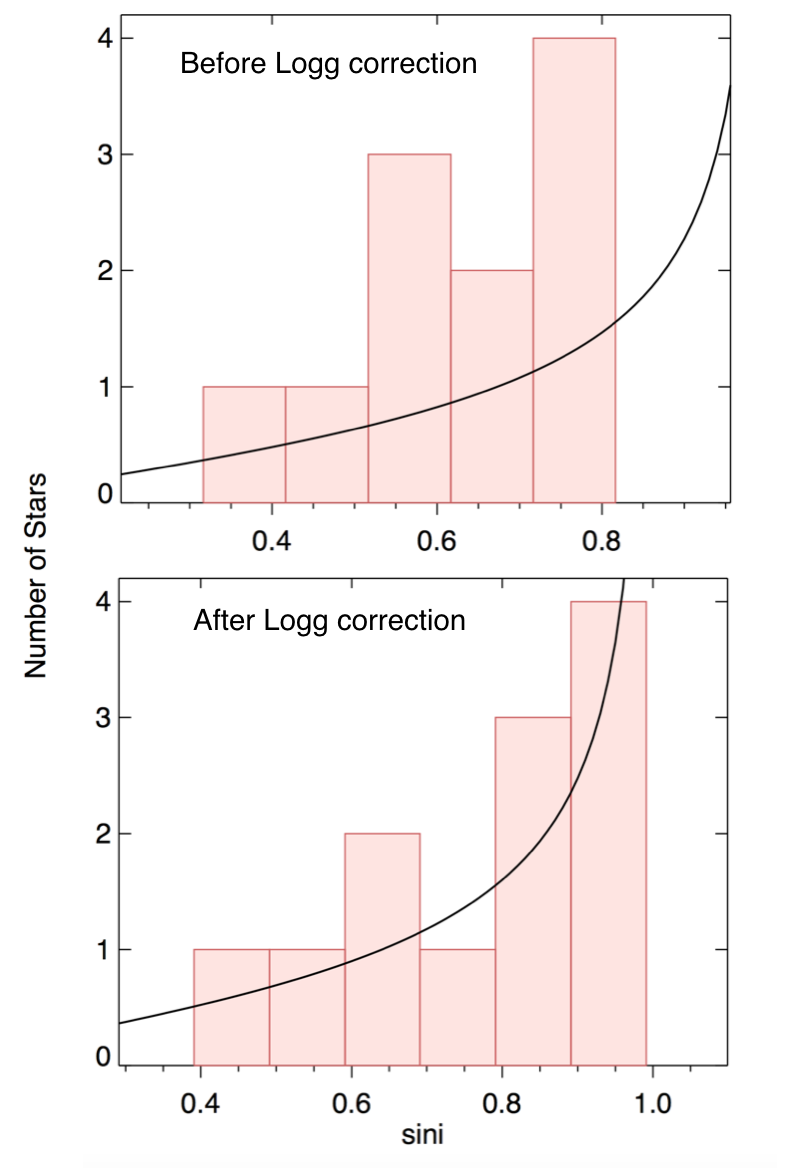}
\caption{Histogram distribution of the number of stars as a function of sini (inclination angle) before (top) and after (bottom) the surface gravity correction for 32 rapid-rotating stars. The black line shows the expected trend of the histogram for a simulated randomly distributed inclination angles for a sample of the same size. The observed distribution is skewed to small angles, peaking at sini= 0.4}
\label{fig:sinihist}
\end{center}
\end{figure}

Since rotation is not being fit directly in the giant regime, the APOGEE ASPCAP pipeline is instead trying to fit the broader lines by incorrectly increasing the surface gravity. Thus, a fraction of rotating giants are likely to have poor-quality spectra fits which potentially contributes to their offsets from the [C/N]-mass relation. To investigate whether this is generally true for the APOKASC sample, we used star-spot modulation, periods, measured vsinis, and asteroseismic radii to calculate rotation velocities, which combined with vsini can inform us about the line-of-sight inclination of each star.%each star's inclination to our line of sight.
 In a field population, we expect the inclination angles to be distributed isotropically  \citep[see e.g.][]{Ceillier2017}. For our sample, we observe a signficantly skewed distribution of inclination angles, peaking around sin(i)=0.7 (Figure \ref{fig:sinihist}, top). This is consistent with overestimated surface gravities causing rotation velocities to be underestimated.

\begin{figure}
\begin{center}
\includegraphics[width=8.5cm]{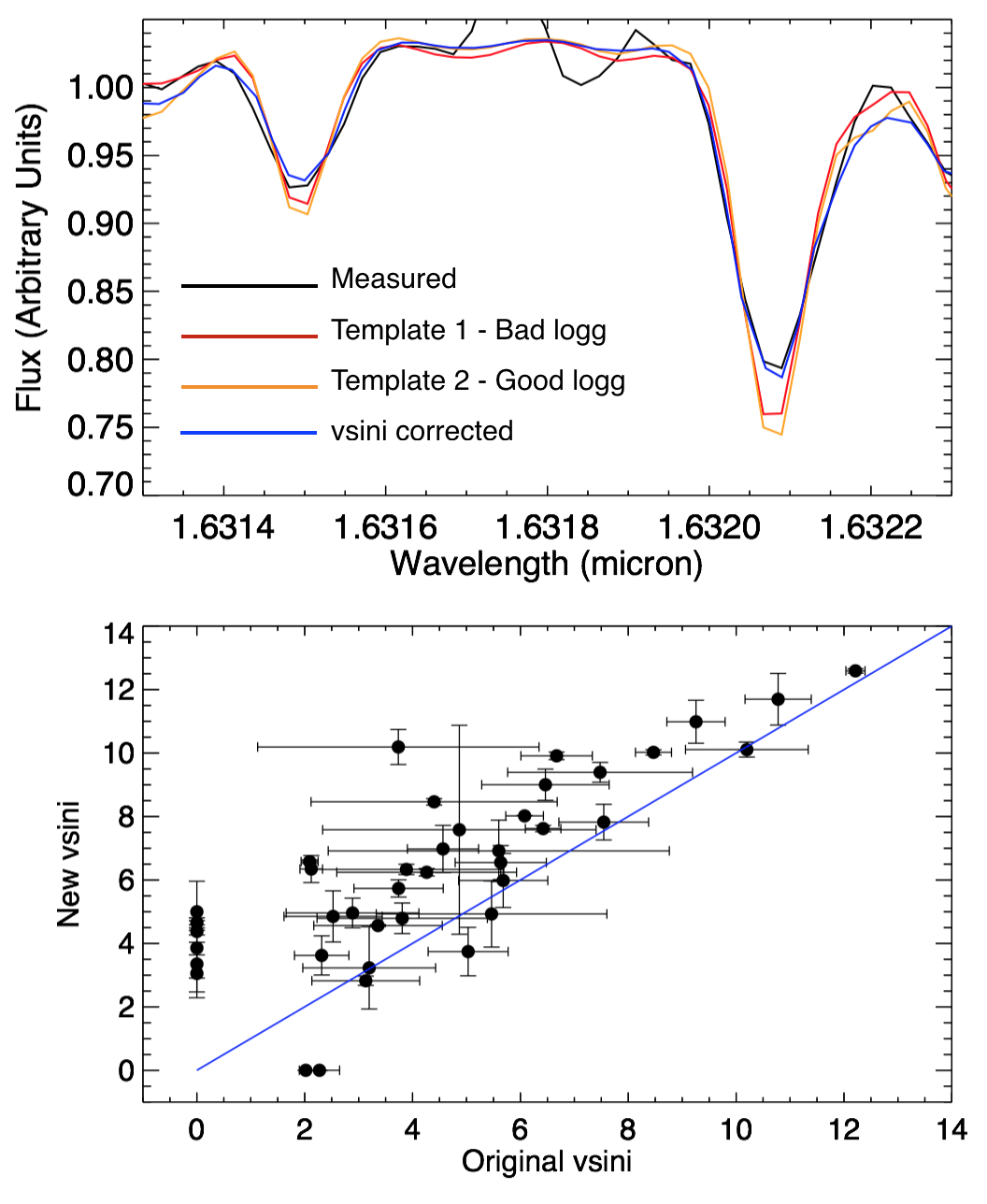}
\caption{Top: Flux versus Wavelength for an individual star observed by APOGEE for a portion of its spectral region. In black is the measured spectrum, red shows the spectral line template by ASPCAP with an incorrect surface gravity parameter (Template 1), orange shows the the same template with a corrected surface gravity using the logg measurement from Kepler (Template 2), and blue shows the line fit after using Template 2 and correcting for rotational broadening (vsini). Bottom: Corrected vsini versus original vsini for stars flagged to have different loggs measured by APOGEE and Kepler. The blue line shows a 1:1 correspondence. Vsinis for our sample are higher on average by 1.7 \kms. 
}
\label{fig:spectra_compare}
\end{center}
\end{figure}

To fix our under-estimated vsin(i) values we adopt the asteroseismic log(g) as the true surface gravity and match each star that has an offset APOGEE logg with a ASPCAP star template. The template is associated with a star that is not rotating detectably but has a surface gravity that matches the asteroseismic gravity of our target as well as its temperature, metallicity, and $\alpha$-element abundance (Figure \ref{fig:spectra_compare}). Next, we recalculated the surface rotation velocity as described in Section %\ref{sec:vsini}
 3.4.2, and compared that to our original measurement. We do this for RC stars that deviate by at least 1 sigma from the APOGEE log(g) vs Kepler log(g) trend (Figure 7, middle panel), which includes both stars from our outlier sample and the parent sample (stars that follow the [C/N]-mass trend).

As expected, we find that the \ensuremath{v \sin{i}}s calculated with the correct surface gravity are systematically higher, by 45\% on average (Figure \ref{fig:spectra_compare}) and that the inferred inclination angles are now consistent with being randomly distributed (Figure \ref{fig:sinihist}, bottom). Furthermore, we identify 6 new stars with measurable vsinis from spectra that were originally hidden by the over-broadened lines in the template. Of these new RC rotators, all but one are stars that fit the [C/N]-mass trend. The new vsinis for stars the outlier sample with overestimated APOGEE log(g)s are in Table (\ref{tab:rotation}). We found that 62\% of our stars with measured rotation needed log(g) and vsini correction. Overall we estimate that about 20\% of outliers had over-estimated log(g)s in comparison to only 5\% of stars in the parent sample, suggesting that [C/N] outliers and rotation are correlated.

\subsection{Stellar Interactions}

\label{massgain}
\begin{figure*}
\begin{center}
\includegraphics[width=18.1cm]{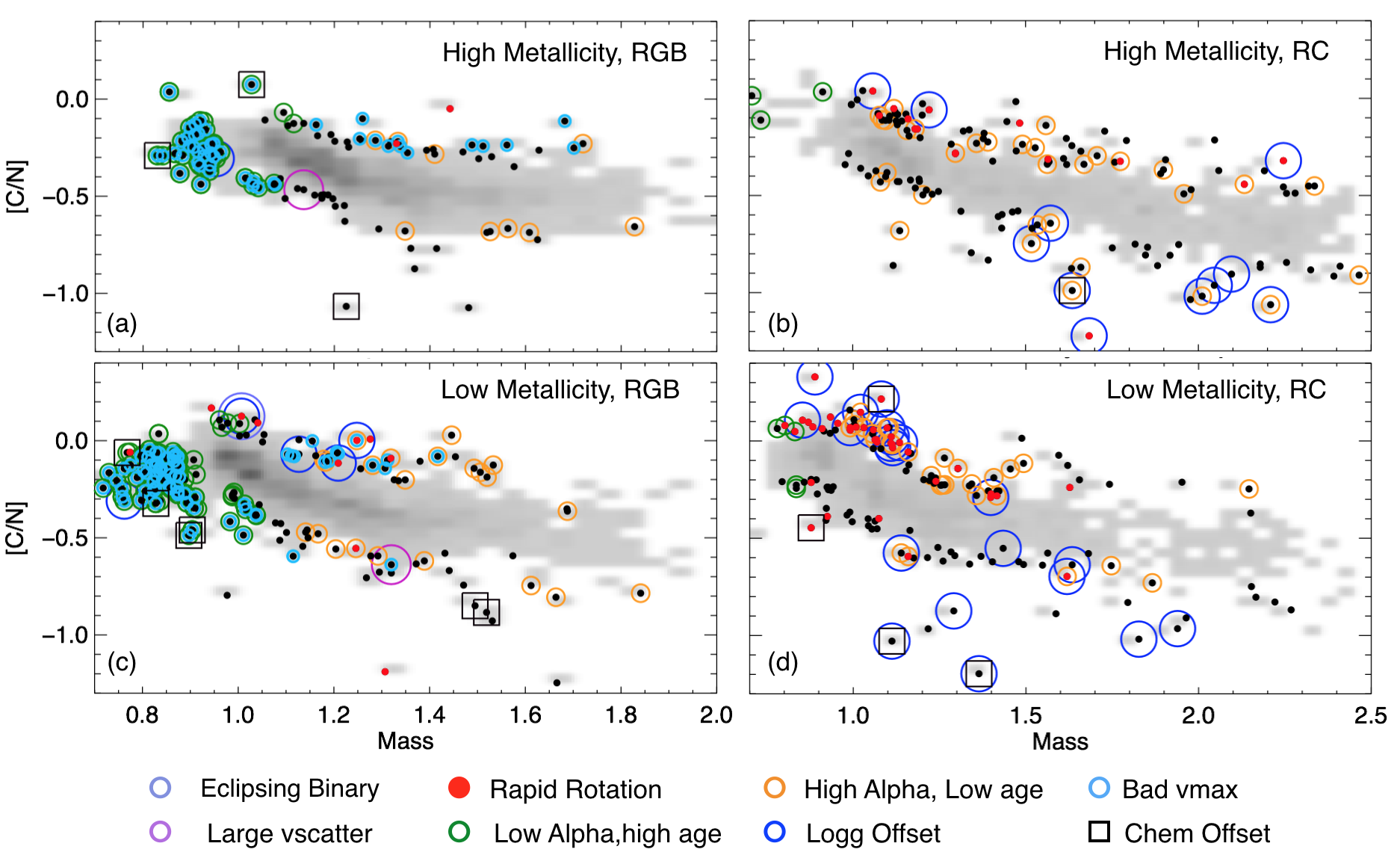}
\caption{A summary of the additional anomalies identified in our outliers from the [C/N]-mass relation for high-metallicity (top) and low-metallicity (bottom) stars as well as first ascent red giants (left) and red clump stars (right). }
\label{fig:lowRC}
\end{center}
\end{figure*}

The main categorizations of outliers are summarized in Figure 10. We compared the characteristics of our outlier sample to the non-outlier sample (i.e. the stars that do follow the [C/N]-mass trend) in Table \ref{tab:properties}. For the low-metallicity red clump population (Figure \ref{fig:lowRC}d), we find a significant number of outliers from the [C/N]-mass relation also have rapid rotation, disagreements between their seismic and spectroscopic surface gravities, and ages that are younger than their $\alpha$-element abundances would suggest. These properties are more prominent in the outlier sample in comparison to the non-outlier sample (Table \ref{tab:properties}, bold items). Many of these are consistent with binary evolution, which is consistent with close binaries being more common in low metallicity stars \citep{Badenes2018, Moe2019} and that red clump stars have already passed through the tip of the red giant branch where their large size would have increased the cross-section for interaction. 

For the high-metallicity red clump population (Figure \ref{fig:lowRC}) we find a still significant number of outliers with $\alpha$-element abundances higher than their age would suggest, but this is comparable to the non-outlier population, and there are fewer stars showing rapid rotation, which could be consistent with suggestions that the binary mass ratio and semi-major axis distribution depends on metallicity \citep{MoeDiStefano2017}. We also note that the significant excess of stars above and to the right of the normal [C/N]-mass relationship could be the result of the accretion of unprocessed material, which would increase a star's mass while simultaneously decreasing its carbon-to-nitrogen ratio \textbf{\citep[see, for comparison, simulations by e.g.][] {Izzard2018}}. 

We show the results for the low-metallicity red giant branch population in Figure \ref{fig:lowRC}c. In this case, outliers are dominated by low-mass stars with low \numax\ values and ages that are much older than would be expected given their $\alpha$-element abundances. These stars account for 50\% of outliers but only 5\% of total RGB population (Table \ref{tab:properties}). Given the challenges of extending asteroseismic techniques to the low-\numax\   regime \citep[see e.g.][]{Stello2014, Pinsonneault2018}, we suggest that these could be stars whose seismic analysis needs to be treated more carefully than the ensemble analysis of \citet{Pinsonneault2018} could allow. On the other hand, if their seismic parameters are reliable, these could be extremely interesting objects for further study of binary stellar evolution, since their large sizes would have increased their interaction cross-sections, and their low masses could suggest that material from the envelope has been lost\textbf{, an interaction that may be expected theoretically to result in lower [C/N] abundances \citep{Izzard2018}}. If these old, low-metallicity stars do turn out to survive more careful asteroseismic analyses, it would also be interesting to determine whether any of them have chemical anomalies and kinematic properties consistent with formation in another galaxy and accretion into the Milky Way \citep[e.g.][]{Grunblatt2021}.

Finally, we show in Figure \ref{fig:lowRC}a the results  for high-metallicity red giant branch stars. In this population, we have a smaller population of outliers 10\% versus 7\% in the low metallicity red clump. As in the low-metallicity red giant branch, the majority of outliers are low-mass, old giants identified as low \numax, which may be the result either of asteroseismic measurement error, or mass loss. 

Lastly, we find that eclipsing binaries or stars with large vscatter do not make up a large fraction of any of our outlier populations. This is consistent with previous results that suggested that the APOKASC analysis preferentially excluded binary systems \citep{Tayar2015}, likely due to the complexity of analysis and decrease in oscillation amplitude in some of these systems\citep{Gaulme2014}. We do detect two new binary systems from high vscatter, KIC~5446355 and KIC~8127707. 

There are rarely conclusive signs that any individual star has undergone an interaction with another star or planet. However, from our analyses we find that most outliers have a signature that is more consistent with binary interactions than internal mixing processes. Using the different diagnostics that we discussed in Section \ref{sec:outlierdiagnostics}, we find strong evidence that many of the outliers from the [C/N]-mass relation are the result of stellar interactions.

\subsection{[C/N] correlation with Iron Peak Elements}

The results in the previous section suggest that many outliers in the [C/N] - stellar mass relation may have gained mass from an interaction with a companion. We therefore search for other unusual abundances which could indicate the source of the accreted mass.  If the source of the additional material is an unevolved low-mass star or substellar companion, then we do not expect offsets in other elements available in the infrared (APOGEE), although optical measurements of lithium abundance could be informative \citep[e.g.][]{aguilera2016, Soares-Furtado2020}. 
On the other hand, material gained from a AGB companion is likely to be enhanced in s-process elements, with an increased total of carbon and nitrogen \citep{Han1995} and pollution from a Type Ia supernova explosion may enhance the abundances of iron-peak elements such as nickel at the stellar surface \citep{GonzalezHernandez2009}.

\begin{figure}
\begin{center}
\includegraphics[width=8.5cm]{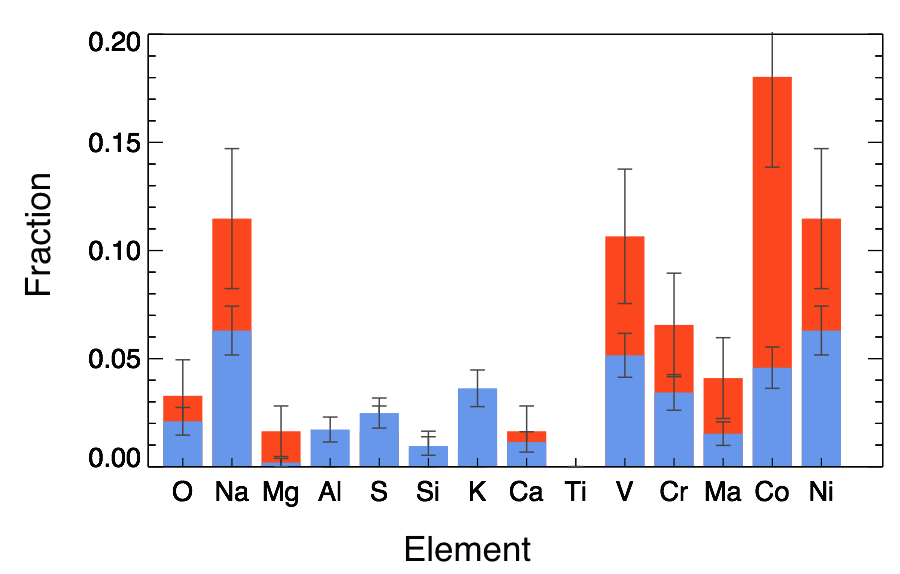}
\caption{
We compare the fraction of stars anomalous in each element for our outlier sample (red) as well as the stars that fall within the expected [C/N]-mass relationship (blue). Our outlier stars are significantly more likely to have e.g. cobalt anomalies than the stars that fit the trend, but it is not clear whether this is due to surface pollution or correlated measurement errors.} 
\label{fig:chemAnom}
\end{center}
\end{figure}

Figure \ref{fig:lowRC} shows stars that are offset from similar stars by more than 0.5 dex in O, Mg, Al, Si, S, K, Ca, Ti, V, Mn, Fe, Co, and Ni. using the procedure discussed in Section \ref{ssec:chemAnom}. Approximately 20-30\% of outliers have chemical offsets in comparison to 5-10\% of non-outlier stars. In Figure \ref{fig:chemAnom} (right) we compare the fraction of stars offset in each element between stars that follow the standard relationship between [C/N] and mass (blue) and those we identify as outliers (red). 

\begin{figure}
\begin{center}
\includegraphics[width=8.5cm]{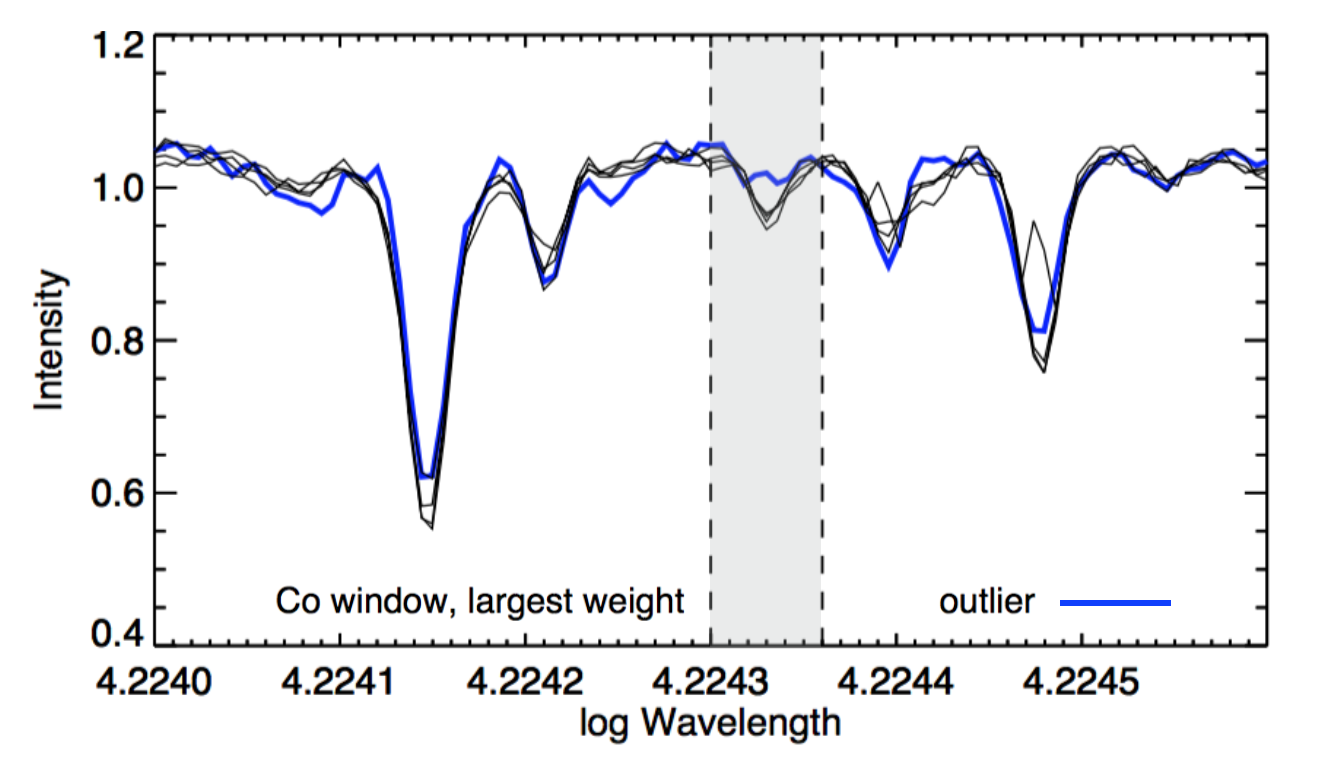}
\caption{For one of our stars KIC 7553192 with an anomalous cobalt abundance, we show a small region around the most highly weighted cobalt line (grey band) and show that the outlier star (blue) has significantly shallower absorption in that region than the four spectroscopically similar stars with standard cobalt abundance (grey lines)}
\label{fig:cobalt}
\end{center}
\end{figure}

The most significant chemical offset is observed for cobalt abundances. We show in Figure \ref{fig:cobalt} one example of a star whose cobalt is significantly lower than stars similar metallicity, $\alpha$ abundance, temperature, and gravity. However, we caution that the wavelength windows used to infer the cobalt abundance in the ASPCAP pipeline have significant overlap with the windows used to derive carbon and nitrogen. Therefore, it is possible that the cobalt anomalies in stars with anomalous carbon and nitrogen abundances could be due to correlated measurement errors between abundances within the ASPCAP pipeline. More generally, it is also possible that some offset in one or more of the bulk parameters of the star such as temperature or surface gravity due to e.g. enhanced rotation may propagate into offsets in all of the chemical abundances that are subsequently fit using the assumed (erroneous) parameter.

A line-by-line analysis, opposed to the spectral synthesis used by the ASPCAP pipeline, could clarify whether these offsets in cobalt, as well as those in nickel, chromium, and manganese, which have similar issues, are real. We also note a slight excess of stars whose sodium abundance is slightly offset from expectations. However, we caution that there are only two spectral windows being used by the ASPCAP pipeline to compute the sodium abundance, and that both are close to telluric features in the spectrum, which suggests that this may be an analysis issue rather than a true physical offset. More generally, we find that none of our stars show conclusive evidence of pollution by a particular source, such as an AGB star or a nearby supernova, although we cannot rule out the possibility that more careful and exhaustive searches for abundance anomalies may clarify the sources of the mass gained in some cases. 

\section{Conclusions}

Galactic Archaeology depends on precise measurements of stellar ages, which requires % To derive a precise age one needs 
a robust measurement of stellar mass. In this paper, we have investigated outliers in the [C/N] - stellar mass relation, which is a promising tool to measure ages for large stellar populations. Our main conclusions are as follows: 

\begin{itemize}

    \item $\approx 10$ \% of red clump stars and first ascent giants do not follow the expected relationship between the carbon-to-nitrogen ratio and mass; there are more deviations at lower metallicities and in the red clump. 
    
    \item For low-metallicity red clump stars that are outliers to the [C/N]-mass trend, we tend to underpredict their mass and thus overpredict their age. For first ascent giants that are outliers to the [C/N]-mass trend, we tend to overpredict the mass and thus underpredict the age.

    \item Stars on the upper giant branch (log(g)$<1$)  seem to deviate from the expected relationship, although we caution that this could be issues with measuring global asteroseismic parameters for evolved stars or known issues for the spectroscopic parameters of the coolest most luminous giants in Data Release 14 \citep{Jonsson2020} rather than a physical deviation from the [C/N]-mass relation. 

    \item Many of the stars that deviate from the [C/N]-mass relation have other properties that are strongly suggestive of having undergone some sort of interaction with another star or sub-stellar companion. Stars with indications of current binarity or past interactions, including rapid rotation, activity, radial velocity variability, and chemical anomalies, are less likely to follow the expected relationship between [C/N] and mass and should be excluded from galactic population studies using [C/N] to estimate masses and ages

    \item We also note that rapidly rotating red clump stars in the APOKASC sample tend to have significantly overestimated spectroscopic surface gravities, affecting 20\% of our outliers and 5\% of the non-outlier sample. Although we have not demonstrated it here, previous work indicates that this may also be correlated with errors in the other stellar parameters measured spectroscopically \citep{Dixon2020,Patton2023}. 
    
\end{itemize}

This is an exciting time for galactic archaeology studies, because the combination of precise asteroseismic data and large spectroscopic surveys will continue to provide the opportunity %not only 
to estimate the masses and ages of large numbers of stars across the galaxy. Machine learning techniques provide a useful tool for extending what can be computed for small samples of stars to estimate the ages of large samples across the galaxy \citep{Ness2016,Mackereth2019, hon2021}. However, %we caution that
as shown here these tools should be applied carefully, as stars that deviate from the normal trend can bias the inferred properties if not taken into  account, and we provide some guidance here on how to exclude those stars. 

%Separately, we note that 
The pathways and results of binary stellar evolution is one of the largest uncertainties in our understanding of low-mass stars and understanding outliers in [C/N]-stellar mass relation. Phases that involve mass transfer and common envelopes are particularly hard to model from first principles. %We suggest that the combination 
Future combinations of spectroscopic, photometric, and asteroseismic diagnostics may provide a reliable way to identify stars that have undergone a recent interaction, and %further study of them could allow us to 
thus provide more robust tools to better constrain the physics of this important phase.

\begin{acknowledgements}

We thank D. Schneider for pointing out helpful references. We thank the anonymous referee for helpful comments that improved this manuscript. 
E.B. and D.H. acknowledge support from the  National Aeronautics and Space Administration (80NSSC19K0597). D.H. also acknowledges support from the Alfred P. Sloan Foundation. Support for this work was provided by NASA through the NASA Hubble Fellowship grant No.51424 awarded by the Space Telescope Science Institute, which is operated by the Association of Universities for Research in Astronomy, Inc., for NASA, under contract NAS5-26555. Funding for the Sloan Digital Sky Survey IV has been provided by
the Alfred P. Sloan Foundation, the U.S. Department of Energy Office of
Science, and the Participating Institutions. SDSS-IV acknowledges
support and resources from the Center for High-Performance Computing at
the University of Utah. The SDSS website is www.sdss.org.

SDSS-IV is managed by the Astrophysical Research Consortium for the 
Participating Institutions of the SDSS Collaboration including the 
Brazilian Participation Group, the Carnegie Institution for Science, 
Carnegie Mellon University, the Chilean Participation Group, the French Participation Group, Harvard-Smithsonian Center for Astrophysics, 
Instituto de Astrof\'isica de Canarias, The Johns Hopkins University, 
Kavli Institute for the Physics and Mathematics of the Universe (IPMU) / 
University of Tokyo, Lawrence Berkeley National Laboratory, 
Leibniz Institut f\"ur Astrophysik Potsdam (AIP),  
Max-Planck-Institut f\"ur Astronomie (MPIA Heidelberg), 
Max-Planck-Institut f\"ur Astrophysik (MPA Garching), 
Max-Planck-Institut f\"ur Extraterrestrische Physik (MPE), 
National Astronomical Observatory of China, New Mexico State University, 
New York University, University of Notre Dame, 
Observat\'ario Nacional / MCTI, The Ohio State University, 
Pennsylvania State University, Shanghai Astronomical Observatory, 
United Kingdom Participation Group,
Universidad Nacional Aut\'onoma de M\'exico, University of Arizona, 
University of Colorado Boulder, University of Oxford, University of Portsmouth, 
University of Utah, University of Virginia, University of Washington, University of Wisconsin, 
Vanderbilt University, and Yale University.

\clearpage 

\startlongtable
\begin{deluxetable*}{ccccccccccccc}
\tablecaption{\label{tab:rotation} APOKASC-2 outlier stars with measured rotation periods}
\vspace{-0.25cm}
\tablehead{
\colhead{Kepler ID}      &
\colhead{$\nu_{max}$}    &
\colhead{$\Delta_{\nu}$} &
\colhead{T$_{eff}$}      &
\colhead{logg (A)}     &
\colhead{logg (K)}     &
\colhead{Mass}   &
\colhead{P$_{rot}$}      &
\colhead{$\sigma_{P_{rot}}$}          &
\colhead{P$_{rot}$ (C)}  &
\colhead{P$_{rot}$ (G)}  &
%%%%%\colhead{P$_{crit}$}     &
\colhead{vsini (old)}  &
\colhead{vsini (new)}   \\
\colhead{}      &
\colhead{}    &
\colhead{} &
\colhead{(K)}      &
\colhead{}     &
\colhead{}     &
\colhead{(M$_{\odot}$)}   &
\colhead{(days)}      &
\colhead{(days)}          &
\colhead{(days)}  &
\colhead{(days)}  &
%%%%%\colhead{P$_{crit}$}     &
\colhead{(km/s)}  &
\colhead{(km/s)}
}
\startdata
2305930 & 28.62 & 3.923 & 4858 & 2.910 & 2.369 & 0.889 & 33.23 & 2.631 & 33.75 & 33.0 & 12.2 & 12.6\\
3937075 & 29.49 & 3.786 & 4724 & 2.492 & 2.376 & 1.113 & 49.54 & 1.669 & - & - & 3.62 & - \\ 
3937217 & 28.99 & 3.746 & 4759 & 2.522 & 2.370 & 1.110 & 57.10 & 2.322 & 54.28 & 57.0 & 6.46 & 9.00\\
4157276 & 39.59 & 4.336 & 4913 & 2.676 & 2.513 & 1.619 & 51.48 & 4.397 & 49.76 & - & 2.10 & 6.58\\
5002776 & 29.45 & 3.741 & 4648 & 2.622 & 2.372 & 1.134 & 76.39 & 4.563 & 72.35 & - & 3.81 & 4.79\\
5879112 & 28.41 & 4.000 & 4932 & 2.550 & 2.369 & 0.853 & 74.19 & 5.110 & - & - & 3.20 & 3.23\\
5961015 & 44.87 & 4.330 & 4939 & 3.040 & 2.568 & 2.246 & 81.69 & 3.425 & - & - & 4.87 & 7.58\\
6289527 & 24.86 & 3.486 & 4810 & 2.350 & 2.306 & 0.910 & 61.39 & 2.045 & - & - & 2.56 & - \\ 
6593240 & 31.16 & 4.245 & 4805 & 2.454 & 2.404 & 0.831 & 103.2 & 5.929 & - & - & 2.14 & - \\ 
6791309 & 28.68 & 3.718 & 4737 & 2.411 & 2.365 & 1.069 & 48.21 & 4.139 & 47.08 & - & 3.79 & - \\ 
7201600 & 29.05 & 3.864 & 4826 & 2.866 & 2.374 & 1.021 & 37.95 & 4.712 & 37.71 & - & 10.78 & 11.70 \\
7205397 & 28.90 & 3.677 & 4631 & 2.537 & 2.363 & 1.114 & 52.99 & 2.017 & 52.6 & - & 3.36 & 4.56 \\
7612916 & 29.22 & 3.887 & 4810 & 2.409 & 2.376 & 0.955 & 94.78 & 6.408 & 102.32 & - & 2.24 & - \\ 
7818452 & 36.32 & 4.187 & 4858 & 2.583 & 2.473 & 1.398 & 32.49 & 4.460 & - & - & 5.60 & 6.92\\
8480313 & 28.16 & 3.647 & 4714 & 2.430 & 2.356 & 1.063 & 60.64 & 12.67 & 60 & - & 4.34 & - \\ 
8879518 & 45.75 & 4.727 & 4839 & 2.772 & 2.572 & 1.684 & 108.9 & 9.522 & 117.57 & - & 2.56 & 3.62 \\ 
8941031 & 28.19 & 3.600 & 4659 & 2.684 & 2.354 & 1.109 & 54.02 & 3.331 & 53.33 & 54.2 & 6.08 & 8.01\\
9086653 & 30.50 & 3.674 & 4603 & 2.702 & 2.385 & 1.221 & 61.01 & 2.415 & 56.76 & - & 5.64 & 6.55\\
9287916 & 31.84 & 3.816 & 4564 & 2.439 & 2.402 & 1.180 & 102.1 & 7.571 & - & - & 2.63 & - \\ 
9833651 & 38.24 & 4.238 & 4680 & 2.521 & 2.487 & 1.483 & 100.0 & 7.981 & 94.15 & - & 2.57 & -\\ 
10482521 & 31.14 & 3.735 & 4694 & 2.333 & 2.399 & 1.239 & 87.77 & 9.314 & - & - & 2.02 & - \\ 
10489017 & 36.93 & 4.267 & 4755 & 2.488 & 2.475 & 1.297 & 92.65 & 6.154 & - & - & 2.31 & - \\ 
10783198 & 27.19 & 3.674 & 4818 & 2.762 & 2.345 & 0.991 & 44.27 & 1.369 & 42.72 & - & 10.2 & 10.11\\
11597759 & 29.10 & 3.748 & 4717 & 2.842 & 2.370 & 1.081 & 47.43 & 1.716 & 46.11 & - & 9.26 & 10.98\\
11808481 & 27.93 & 3.774 & 4814 & 2.453 & 2.357 & 0.934 & 54.46 & 1.985 & 52.23 & - & 5.91 & - \\ 
11913049 & 28.58 & 3.664 & 4605 & 2.662 & 2.357 & 1.098 & 48.21 & 1.620 & 48.74 & - & 5.47 & 4.93 \\
12011705 & 29.82 & 4.150 & 5006 & 2.356 & 2.394 & 0.878 & 91.90 & 6.763 & - & - & 2.10 & 0.00 \\
12204548 & 28.94 & 3.679 & 4763 & 2.396 & 2.370 & 1.187 & 62.54 & 2.191 & - & - & 2.43 & 0.00 \\
12304485 & 28.69 & 3.773 & 4821 & 2.514 & 2.369 & 1.058 & 53.96 & 3.095 & 53.33 & - & 7.48 & 9.39 \\
\hline
\enddata
\vspace{0.2cm}
{\bf Notes:} : P$_{rot}$, P$_{rot}$ (C) and P$_{rot}$ (G) refer to rotation period from this paper, Cellier et. al (2017) and Gaulme et al. (2020) respectively. Old vsini and new vsini refer to vsini from this paper before and after logg correction (see section 4.1).
\vspace{3cm}
\end{deluxetable*}

\startlongtable
\begin{deluxetable*}{llcc}
\tablecaption{\label{tab:properties} Fractions of outlier Stars with different characteristics in comparison to stars that follow the CN-mass trend.}
\vspace{-0.25cm}
\tablehead{
\colhead{Population Type}      &
\colhead{Category}    &
\colhead{Outlier Fraction} &
\colhead{Non-Outlier Fraction} }
\startdata
RC - Low [Fe/H] & known binaries$^{a}$ & 0.037 & 0.049 \\
& high alpha, low age & { 0.224} & 0.168 \\
& low alpha, high age & { 0.045} & 0.018 \\
& Low numax & 0 & 0 \\
& vscatter $>1$ & 0 & 0 \\
& vsini$^{b}$ $>2$ & { 0.284 (0.291)} & 0.113 (0.118) \\
& bad logg & { 0.157} & 0.039 \\
& chemical anomaly$^{c}$ & { 0.366} & 0.138 \\
\hline
RC - High [Fe/H] & known binaries & 0.059 & 0.069 \\
& high alpha, low age & 0.237 & 0.261 \\
& low alpha, high age & 0.022 & 0.019 \\
& Low numax & 0 & 0 \\
& vscatter $>1$ & 0 & 0.005 \\
& vsini $>2$ & {0.111} & 0.031 (0.032) \\
& bad logg & { 0.081} & 0.018 \\
& chemical anomaly & { 0.363} & 0.109 \\
\hline
RGB - Low [Fe/H] & known binaries & 0.030 & 0.064 \\
& high alpha, low age & 0.132 & 0.141 \\
& low alpha, high age & { 0.669} & 0.257 \\
& Low numax & {0.452} & 0.073 \\
& vscatter $>1$ & 0.006 & 0.01 \\
& vsini $>2$ & { 0.060} & 0.027 \\
& bad logg & {0.030} & 0.005 \\
& chemical anomaly & { 0.193} & 0.073 \\
\hline
RGB - High [Fe/H] & known binaries & 0.087 & 0.081 \\
& high alpha, low age & 0.087 & 0.160 \\
& low alpha, high age & { 0.683} & 0.159 \\
& Low numax & { 0.471} & 0.070 \\
& vscatter $>1$ & 0.010 & 0.010 \\
& vsini $>2$ & { 0.019} & 0.004 \\
& bad logg & { 0.01}  & 0.003 \\
& chemical anomaly & { 0.288} & 0.055 \\
\hline
\enddata
%\tablenotes
\vspace{0.2cm}
{\bf Notes:} $^{a}$ Known Binaries includes those flagged as eclipsing binaries, RUWE $>1.2$, or part of the Price-Whelan sample. \\
$^{b}$ In parenthesis is the updated fraction after logg correction where applicable. \\
$^{c}$ Fraction of targets that have one or more chemical abundances that deviate from 4 stars with similar stellar properties that follow the CN-mass trend.
\end{deluxetable*}

\end{acknowledgements}

\bibliographystyle{aasjournal}
\bibliography{ms}  

\end{document}